\documentclass[12pt]{article}

\usepackage{amsmath, amssymb, amsfonts, multirow}
\usepackage[UKenglish]{babel}
\usepackage[T1]{fontenc}
\usepackage[utf8]{inputenc}	
\usepackage{graphicx}
\usepackage{multirow}
\usepackage{hyperref}
\usepackage{braket}
\usepackage{indentfirst}
\usepackage{float}
\usepackage{caption}
\usepackage{subcaption}

\usepackage[numbered,framed]{matlab-prettifier}

\usepackage{authblk} 

\usepackage{xcolor}

\begin{document}

\title{Efficient numerical method to handle boundary conditions in 2D elastic media}
\author[1]{Dénes Berta}
\author[1]{István Groma}
\author[1]{Péter Dusán Ispánovity}
\affil[1]{Eötvös Loránd University, Department of Materials Physics, 1117 Budapest, Pázmány Péter Sétány 1/a. Hungary}


\frenchspacing

\maketitle
\thispagestyle{empty}




\hspace{0pt}
\vfill

\begin{abstract}
A numerical method is developed to efficiently calculate the stress (and displacement) field in finite 2D rectangular media. The solution is expanded on a function basis with elements that satisfy the Navier--Cauchy equation. The obtained solution approximates the boundary conditions with their finite Fourier series. The method is capable to handle Dirichlet, Neumann and mixed boundary value problems as well and it was found to converge exponentially fast to the analytical solution with respect to the size of the basis. Possible application in discrete dislocation dynamics simulations is discussed and compared to the widely used finite element methods: it was found that the new method is superior in terms of computational complexity.
\end{abstract}
\vfill
\hspace{0pt}

\newpage

\section{Introduction}

Several mechanical properties of crystalline matter, such as work hardening, ductile-brittle transition, creep or fatigue are caused by the collective motion of lattice dislocations \cite{hull2001introduction}. These defects, therefore, have played a central role in materials science in the last approx.~80 years. In order to describe the mentioned and related phenomena one has to understand both the individual properties of dislocations (usually investigated using molecular dynamics simulations or methods derived from first principles) and also their complex collective dynamics during plastic flow. The latter is usually investigated on various scales \cite{bulatov2006computer}:
\begin{itemize}
\item The basic constituent in \emph{molecular dynamics (MD) simulations} is the atom, and dislocations form and move in such models as realistic topological crystal defects . Although due to the huge degrees of freedom in such models the simulations are strongly constrained both in achievable volume (typically less then 1 $\mu$m$^3$) and duration (typically few ps), this method is very powerful since it gives the best possible description of dislocation dynamics without any significant approximation \cite{wolf2005deformation, domain2005simulation, zepeda2017probing, niiyama2015atomistic}.
\item In the case of \emph{discrete dislocation dynamics (DDD)} the basic constituents are the dislocation lines themselves and the underlying crystal lattice is considered as a continuum elastic medium. During the simulations complicated dynamic equations govern the motion and interaction of dislocations. These equations are either derived using physical arguments or by lower scale numerical modelling \cite{van1995discrete2, cleveringa1997comparison, benzerga2003incorporating, yefimov2004comparison, bulatov2006dislocation, el2008self, fivel2008discrete, zhou2010discrete, song2019discrete}.
\item In \emph{continuum crystal plasticity (CCP) models} even the dislocations are considered in a continuum manner in the form of various dislocation density fields. The evolution of these fields, and thus the plastic response, is obtained by solving partial differential equations describing the evolution of the densities \cite{el2000statistical, arsenlis2002modeling, yefimov2004comparison, chen2010bending, hochrainer2014continuum, groma2015scale, xia2015computational, hochrainer2016thermodynamically, groma2016dislocation}.
\end{itemize}
All these methods have their advantages and disadvantages and usually one has to choose the one that suits the problem at hand the best.

In the last decades new technologies have emerged to create and manipulate samples on the micron or sub-micron scale. It turned out that at this scale the mechanical properties of crystalline materials are profoundly different from those of macroscopic samples. First of all, a significant size effect can be observed, i.e, the strength of the specimens increases as the size at least in one dimension reduces to or below approx.~10 $\mu$m \cite{fleck1994strain, uchic2009plasticity}. In addition, the plastic response becomes jerky and unpredictable as random strain bursts start to dominate the deformation. These bursts are localized both in time and space and are caused by the sudden rearrangement of the dislocation network \cite{weiss2003three, dimiduk2006scale, csikor2007dislocation}.

These two important features observed experimentally at the smallest scales drew significant attention from the modelling community with the motivation to develop a detailed physical understanding of these phenomena. It is evident, that at small scales sample boundaries play a crucial role. They modify the stress fields of dislocations within the crystal and, thus, act as attracting or repelling surfaces depending on the type of the boundary (fixed stress or displacement). Since in small specimens a large portion of dislocations is close to the surface, one must take boundary conditions properly into account to give a physically correct description. In MD simulations this can be performed by prescribing displacements or forces on the atoms on the boundaries. In higher scale models (that is, DDD and CCP), however, the crystal is modelled as an elastic medium, so, boundary conditions must be solved in the framework of continuum elasticity. For this purpose the elastostatic equations are typically solved using the finite element method (FEM). This versatile and flexible tool allows us to study various geometries and boundary conditions with high numerical stability. Despite of the advantages of the FEM, in some cases different methods may suit the investigated problem better and may, e.g., exhibit faster runtime compared to FEM. It was shown, for instance, by Wei et al.~that a particular spectral method has superior time complexity compared to FEM when modelling 3D DDD in a cylindrical micropillar geometry \cite{weinberger2007computing}. This method is based on the series expansion of the analytical elastic solution and the boundary conditions are prescribed in terms of Fourier coefficients of the desired boundary values.

In this paper we follow the route proposed by Wei et al.~in order to develop a spectral method to efficiently handle the boundary problem for 2D systems. 2D modelling represents an essential part of the numerical research in the field because the conceptual simplicity compared to 3D systems makes it easier (or even possible) to develop and test analytical models of plastic deformation. Consequently, various phenomena has been investigated using 2D models such as thin film plasticity \cite{ayas2008dislocation, fan2011thickness, davoudi2014bauschinger}, micropillar plasticity \cite{benzerga2009micro, song2019universality} and statistics of strain bursts \cite{ispanovity2010submicron, ispanovity2014avalanches, ovaska2015quenched}. All of these studies consider a rectangular simulation area and apply FEM to tackle the boundary problem. Here we, therefore, aim at developing a spectral method that can solve the elastostatic equations on a 2D rectangular domain more efficiently than FEM.

The paper is organized as follows. Firstly, the theoretical background of the method and its main principle, the superposition method introduced by van der Giessen are reviewed in Sec.~\ref{sec:bckgrnd}. This is followed by the presentation of the basis of functions on which we expand our solution of the Navier—Cauchy equation that describes the elastic, homogeneous and isotropic medium in equilibrium. After that the details of the implementation are summarized (Sec.~\ref{sec:implementation}). In Sec.~\ref{sec:results} the method is tested on analytically solvable problems which yields a remarkably fast convergence to the solution. The method is also tested on systems with discrete dislocations and the results are compared with analytic solutions. In the last part of the section the computational complexity of the method is assessed and we indeed obtain a superior performance compared to FEM. Finally, a discussion and summary conclude the paper.

\section{Theoretical background}
\label{sec:bckgrnd}

\subsection{Boundary conditions in 2D dislocation systems}

The stress at a given point of the 2D material can be decomposed into two parts: one part is due to the dislocations and the other is due to external load. The formulae of the stress field of straight dislocations are well-known in an infinite elastic medium. However, in real (finite) systems these solutions do not satisfy the prescribed boundary conditions. In addition, external load is applied on the boundaries in the form of traction or displacement, that usually leads to an inhomogeneous stress field in the material. Since the dynamics of dislocations is governed by the local stress via the Peach--Koehler equation
\begin{equation}
\mathbf{F}=\mathbf{l}\times(\sigma\mathbf{b})
\label{eq:peachkoehler}
\end{equation}
(where $\mathbf{F}$ is the force acting on the unit length of a dislocation line, $\mathbf{l}$ is the unit vector pointing in the direction of the dislocation line, $\mathbf{b}$ is the Burgers vector, and $\sigma$ is the stress tensor at the position of the dislocation), the boundaries may significantly affect the acting forces and, thus, the evolution of dislocation ensembles. Hence, it is very important to handle properly the boundary conditions in DDD simulations.

The boundary condition may concern the displacement (Dirichlet boundary value problem), the stress (Neumann boundary value problem) or the displacement on some parts of the boundary and the stress on the others (mixed boundary value problem). The numerical method proposed in this article is capable of solving all of these on a rectangular 2D domain.

\begin{figure}[H]
\centering
\includegraphics[width=11cm]{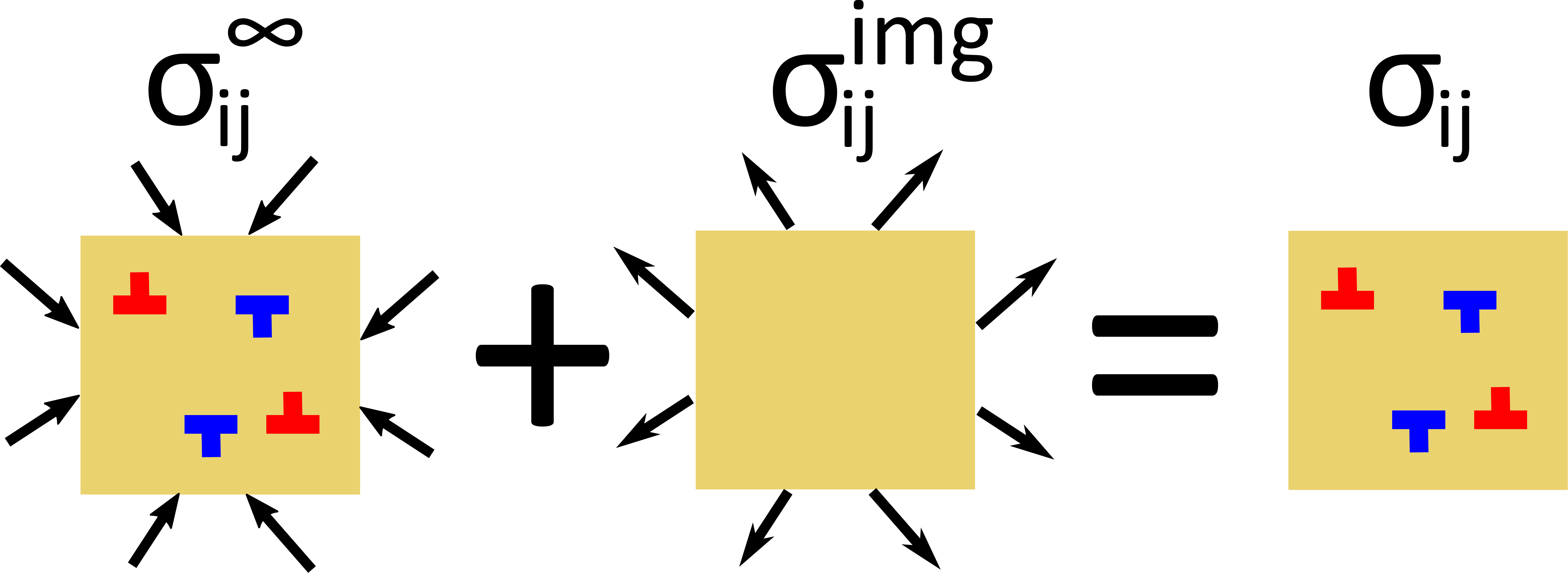}
\caption{Illustration of the practical decomposition of the $\sigma_{ij}$ stress field (equation \ref{eqn:stress_decomposition}): the first term $\sigma_{ij}^\infty$ is the field of dislocations as if they were in an infinite medium and the second one ($\sigma_{ij}^{\mathrm{img}}$) is a dislocationless solution of the inverse elastic problem. The superposition of this two fulfils the equation describing elastic media while containing dislocations and satisfying the boundary conditions as well. In the case shown in the figure this means that the relevant components of the stress vanish on the boundary, although, the method can be applied to solve problems with arbitrary boundary conditions.}
\label{fig:vdg}
\end{figure}

To describe the stress field of dislocations that fulfil the boundary conditions we follow the method proposed by van der Giessen and Needleman \cite{van1995discrete}. Assuming linear elasticity the stress field can be decomposed in the following way:
\begin{equation}
\sigma_{ij}=\sigma_{ij}^\infty+\sigma_{ij}^{\mathrm{img}},
\label{eqn:stress_decomposition}
\end{equation}
where $\sigma_{ij}^\infty$ is the stress field of the dislocations as if they were in an infinite medium and $\sigma_{ij}^{\mathrm{img}}$ is a dislocationless solution of the elastic problem with complementary boundary conditions. The latter is defined so that the superposition of the two must satisfy the boundary conditions determined by the examined physical problem. The method is illustrated in figure \ref{fig:vdg} for a stress-free boundary.

Since one can easily evaluate the stress field of dislocations in infinite medium (that is, $\sigma_{ij}^\infty$), the relevant task is to solve the dislocationless elastic problem in finite medium with given boundary conditions (which may concern displacement, stress or both) in order to obtain $\sigma_{ij}^{\mathrm{img}}$. We, therefore, continue with solving this elastic problem.

\subsection{Solution of Navier-Cauchy equation in 2D}

To obtain the equilibrium displacement field of a homogeneous and isotropic elastic medium one should solve the Navier--Cauchy equation:
\begin{equation}
\mu\Delta\mathbf{u}+(\lambda+\mu)\nabla(\nabla\mathbf{u})=0,
\label{eq:navca}
\end{equation}
where \textbf{u} is the displacement and $\lambda$ and $\mu$ are Lamé's first parameter and the shear modulus, respectively. As it was discussed in the Inroduction in this paper we focus on 2D problems, so, in the following we will solve the Navier--Cauchy equation in 2D, that is, when the solution is invariant in the direction parallel to the $z$ axis.


To solve equation (\ref{eq:navca}) with certain boundary conditions, firstly, we searched for a set of functions that satisfy the equation (without considering any specific boundary conditions). Since, the equation is linear and homogeneous, any linear combination of these functions is a solution as well.
To find the proper basis functions, the displacement field can be decomposed to irrotational and solenoidal fields:
\begin{equation}
\mathbf{u}=\nabla\varphi+\nabla\times\psi,
\end{equation}
which in 2D reads as
\begin{gather}
u_x=\partial_x\varphi+\partial_y\psi,\\
u_y=\partial_y\varphi-\partial_x\psi.
\label{eq:pot}
\end{gather}
Since in this case only the $z$ component of the $\psi$ potential is non-vanishing, we simply wrote $\psi$ instead of $\psi_z$. Plugging it into equation (\ref{eq:navca}) yields the biharmonic equation for the two potentials \cite{timoshenko1969theory}:
\begin{gather}
\Delta^2\varphi=0,\\
\Delta^2\psi=0.
\end{gather}
Searching the potentials (for example $\varphi$) in the form
\begin{equation}
\varphi(x,y)=X(x)Y(y)
\end{equation}
yields:
\begin{equation}
\frac{\partial^4 X}{\partial x^4}+\frac{2}{Y}\frac{\partial^2 X}{\partial x^2}\frac{\partial^2 Y}{\partial y^2}+\frac{X}{Y}\frac{\partial^4 Y}{\partial y^4}=0.
\end{equation}
Since, $\frac{1}{Y}\frac{\partial^2 Y}{\partial y^2}$ and $\frac{1}{Y}\frac{\partial^4 Y}{\partial y^4}$ are independent of $x$ (so, they are constants) the solution can be found in the form
\begin{equation}
Y(y)=a\sin\left(\frac{2\pi}{l}ny\right)+b\cos\left(\frac{2\pi}{l}ny\right),
\end{equation}
where $a$ and $b$ arbitrary parameters of length dimension, $l$ is the longest possible wavelength and $n$ is a dimensionless positive integer. The differential equation for $X$ is then:
\begin{equation}
\frac{l^4}{(2\pi)^4}\frac{\partial^4 X}{\partial x^4}-2n^2\frac{l^2}{(2\pi)^2}\frac{\partial^2 X}{\partial x^2}+n^4 X=0,
\end{equation}
which yields
\begin{equation}
X(x)=(a_0+a_1x)e^{\frac{2\pi}{l}nx}+(b_0+b_1x)e^{-\frac{2\pi}{l}nx}.
\end{equation}
Using the results for $X$ and $Y$ the possible solutions can be \cite{hirth1992theory}:
\begin{equation}
	\begin{gathered}
	\phi_1(x,y)=f(2\pi nx/l)g(2\pi ny/l)\\
	\phi_2(x,y)=g(2\pi nx/l)f(2\pi ny/l)\\
	\phi_3(x,y)=xg(2\pi nx/l)f(2\pi ny/l)\\
	\phi_4(x,y)=f(2\pi nx/l)yg(2\pi ny/l),
	\end{gathered}
	\label{eq:biharmsolution}
\end{equation}
where
\begin{equation}
f(s)=\sin{s}~~~\mathrm{or}~~~f(s)=\cos{s}
\label{eq:f}
\end{equation}
and
\begin{equation}
g(s)=e^{\pm s}.
\label{eq:g}
\end{equation}
 The derivatives of the $\phi_i$s can be given with their linear combinations. Therefore, according to equations (\ref{eq:pot}), $u_x$ and $u_y$ can be also written as the linear combination of these functions.

The functions (\ref{eq:biharmsolution}) all satisfy the biharmonic equation, however, only certain linear combinations fulfil the Navier--Cauchy equation (\ref{eq:navca}). The appropriate linear combinations can be determined by inserting the functions from (\ref{eq:biharmsolution}) into the Navier--Cauchy equation as the $u_x$ or $u_y$ component of the displacement have relations between their coefficients. The obtained basis functions (that satisfy the Navier--Cauchy equation) of $u_x$ and $u_y$ are shown in table \ref{tab:base} in which we introduced $k_0=2\pi/l$.

\begin{table}[H]
\centering
\begin{tabular}{|c|c|c|}
\hline
coefficient & $u_x$                           & $u_y$                           \\ \hline
$C_1^{(n)}$ & $(1+\alpha nk_0 y)\sin(nk_0x)e^{nk_0y}$  & $-\alpha nk_0 y\cos(nk_0x)e^{nk_0y}$     \\ \hline
$C_2^{(n)}$ & $(1-\alpha nk_0 y)\sin(nk_0x)e^{-nk_0y}$ & $-\alpha nk_0 y\cos(nk_0x)e^{-nk_0y}$    \\ \hline
$C_3^{(n)}$ & $(1+\alpha nk_0 y)\cos(nk_0x)e^{nk_0y}$  & $\alpha nk_0 y\sin(nk_0x)e^{nk_0y}$      \\ \hline
$C_4^{(n)}$ & $(1-\alpha nk_0 y)\cos(nk_0x)e^{-nk_0y}$ & $\alpha nk_0 y\sin(nk_0x)e^{-nk_0y}$     \\ \hline
$C_5^{(n)}$ & $(1-\alpha nk_0 x)e^{nk_0x}\sin(nk_0y)$  & $-\alpha nk_0 xe^{nk_0x}\cos(nk_0y)$     \\ \hline
$C_6^{(n)}$ & $(1+\alpha nk_0 x)e^{-nk_0x}\sin(nk_0y)$ & $-\alpha nk_0 xe^{-nk_0x}\cos(nk_0y)$    \\ \hline
$C_7^{(n)}$ & $(1-\alpha nk_0 x)e^{nk_0x}\cos(nk_0y)$  & $\alpha nk_0 xe^{nk_0x}\sin(nk_0y)$      \\ \hline
$C_8^{(n)}$ & $(1+\alpha nk_0 x)e^{-nk_0x}\cos(nk_0y)$ & $\alpha nk_0 xe^{-nk_0x}\sin(nk_0y)$     \\ \hline
$C_{9}^{(n)}$ & $-\alpha nk_0 y\cos(nk_0x)e^{nk_0y}$     & $(1-\alpha nk_0 y)\sin(nk_0x)e^{nk_0y}$  \\ \hline
$C_{10}^{(n)}$ & $-\alpha nk_0 y\cos(nk_0x)e^{-nk_0y}$    & $(1+\alpha nk_0 y)\sin(nk_0x)e^{-nk_0y}$ \\ \hline
$C_{11}^{(n)}$ & $\alpha nk_0 y\sin(nk_0x)e^{nk_0y}$      & $(1-\alpha nk_0 y)\cos(nk_0x)e^{nk_0y}$  \\ \hline
$C_{12}^{(n)}$ & $\alpha nk_0 y\sin(nk_0x)e^{-nk_0y}$     & $(1+\alpha nk_0 y)\cos(nk_0x)e^{-nk_0y}$ \\ \hline
$C_{13}^{(n)}$ & $-\alpha nk_0 xe^{nk_0x}\cos(nk_0y)$     & $(1+\alpha nk_0 x)e^{nk_0x}\sin(nk_0y)$  \\ \hline
$C_{14}^{(n)}$ & $-\alpha nk_0 xe^{-nk_0x}\cos(nk_0y)$    & $(1-\alpha nk_0 x)e^{-nk_0x}\sin(nk_0y)$ \\ \hline
$C_{15}^{(n)}$ & $\alpha nk_0 xe^{nk_0x}\sin(nk_0y)$      & $(1+\alpha nk_0 x)e^{nk_0x}\cos(nk_0y)$  \\ \hline
$C_{16}^{(n)}$ & $\alpha nk_0 xe^{-nk_0x}\sin(nk_0y)$     & $(1-\alpha nk_0 x)e^{-nk_0x}\cos(nk_0y)$ \\ \hline
\end{tabular}
\caption{The basis functions that fulfil equation (\ref{eq:navca}) and the notation for their coefficients, where $\alpha=\frac{\mu+\lambda}{3\mu+\lambda}$ and $k_0=\frac{2\pi}{l}$ .  Apparently, every order of $n$ consists of 16 basis functions and both components are non-vanishing for every function.}
\label{tab:base}
\end{table}

In table \ref{tab:base} the notation
\begin{equation}
\frac{\mu+\lambda}{3\mu+\lambda}=\frac{1}{3-4\nu}\equiv \alpha
\label{eq:alfa}
\end{equation}
is used, where $\nu=\frac{\lambda}{2(\lambda+\mu)}$ is the Poisson ratio. Interestingly, displacement components of the basis functions only depend on one elastic constant instead of two. Mathematically this is the consequence of equation (\ref{eq:navca}) being homogeneous.

The dynamics of dislocations is determined by the stress field, thus, one should calculate the stress field as well. If the displacement field (i.e. the $C_i^{(n)}$ coefficients for $i=1,2,...,16$) is known, the stress field can be easily calculated from it with its derivatives. We determined the stress components corresponding to each basis function, which are not shown explicitly here.

To find the best approximate solution for the elastic problem examined, we should find the linear combination of the basis functions (values for the $C_i^{(n)}$ coefficients) that is the best in some sense. Since all the basis functions in table \ref{tab:base} satisfy the (\ref{eq:navca}) Navier--Cauchy equation, their any linear combination will do as well, however, if we use finite orders of $n$ there might not be a linear solution that fulfils exactly the boundary conditions. In the next section we will show our approach to find a approximate solution that matches the boundary conditions the best.

\section{Implementation to 2D rectangular domain}

\label{sec:implementation}

Using only a finite number of basis functions, the solution might only approximately fulfil the boundary conditions. Our requirement is that the Fourier series of the approximate solution along the boundary and the Fourier series of the boundary condition should be identical in the first finite number of Fourier coefficients. In this section we will introduce the method to determine the solution that meets this criterion.

Supposing that solution (i.e., the finite set of $C_i^{(n)}$ coefficients) is already known, both the stress and displacement can be evaluated at the boundaries. If we settle the coefficients $C_i^{(n)}$ in vector \textbf{c} and the Fourier coefficients of the solution on the boundary in vector \textbf{f}, there is a linear relation between the two. It can be described with matrix \textbf{M} as:
\begin{equation}
\mathbf{f}=\mathbf{M}\mathbf{c}.
\end{equation} 
Since, the boundary condition is known, the task is to determine the $C_i^{(n)}$ coefficients. It can obtained by inverting the matrix \textbf{M}:
\begin{equation}
\mathbf{c}=\mathbf{M}^{-1}\mathbf{f}.
\end{equation}
The matrix inversion requires \textbf{M} to be a square matrix. In principle, we have infinite number of basis functions and the Fourier series of the boundary conditions consist of infinite number of coefficients as well. Thus, both \textbf{c} and \textbf{f} consist of infinite number of components. In order to be able to accomplish the calculation the basis functions and the Fourier coefficients have to be restricted to a finite order. This will result a finite-sized matrix \textbf{M}. If we have basis functions of orders $n=1,2,...,N$ the vector \textbf{c} 
 will have $16N$ components according to table \ref{tab:base}. It will be shown below that it is enough to use only sinusoidal or cosinusoidal modes of the Fourier series of the boundary conditions. Since, on all four boundaries we need the Fourier coefficients of two displacement or stress components (depending on the boundary conditions) the vector \textbf{f} will contain 8 coefficients of each Fourier order. Hence, to get a square matrix we will need to take into consideration the first $2N$ orders of the Fourier series of the boundary conditions.
 
 Since, the displacement and stress fields do not have physical meaning outside the rectangular area of interest (for which $0\leq x,y \leq L$) we have the freedom to have an arbitrary field there. There are two simple ways to extend the field outside the specimen which we implemented: a field that is periodic or antiperiodic in $x$ and $y$ with (anti)period $L$. During implementation, for simplicity, we used $L=l/2=\pi$, hence, $k_0=1$. This choice, however, does not limit the applicability of the method, since the solution can be easily rescaled to any $L$ value. A periodic field makes all sinusoidal Fourier coefficients vanish while an antiperiodic one will cancel all cosinusiodal coefficients. Therefore, it is possible to use purely sinusoidal or cosinusoidal Fourier-series to describe the boundary conditions. This implies that (as it was mentioned above) $2N$ Fourier orders should be used to describe the boundary conditions if we have a basis of order $N$. In the sinusoidal case we should use Fourier coefficients of order $n=1,2,...,2N$ while for the cosinusoidal case relevant orders are $n=0,1,...,2N-1$, since in the latter case all cosines have zero average, therefore, the constant should be also included in the Fourier series. In practice we determined the components of the vector \textbf{f} using FFT (fast Fourier transform) algorithm after completing the field periodically or antiperiodically.
 
 The $M_{ij}$ matrix element describes the connection between the basis function corresponding to the $j^{\mathrm{th}}$ component of the vector \textbf{c} and the Fourier coefficient of a certain displacement or stress component on a certain boundary corresponding to the $i^{\mathrm{th}}$ component of the vector \textbf{f}. Since, we have finite number of basis number families (namely 16) given by their analytical formula, the matrix element can be calculated either numerically or symbolically. The matrix is unchanged during a DDD simulation (and so is its inverse), hence, the matrix is to be evaluated only once.
 
 The vector \textbf{f} and the matrix \textbf{M} can consist of the Fourier coefficients of displacement or stress or both on the boundaries, otherwise the method remains the same, that is, only the actual vector components and matrix elements depend on the boundary conditions.

\section{Results}
\label{sec:results}
\subsection{Tests on analytically solvable examples}
To verify the validity of our method we tested it on several analytically solvable examples. We examined the convergence of the method numerically on these test cases and we observed especially fast convergence as it is demonstrated in detail below.

\begin{figure}[H]
\centering
\includegraphics[width=0.5\textwidth]{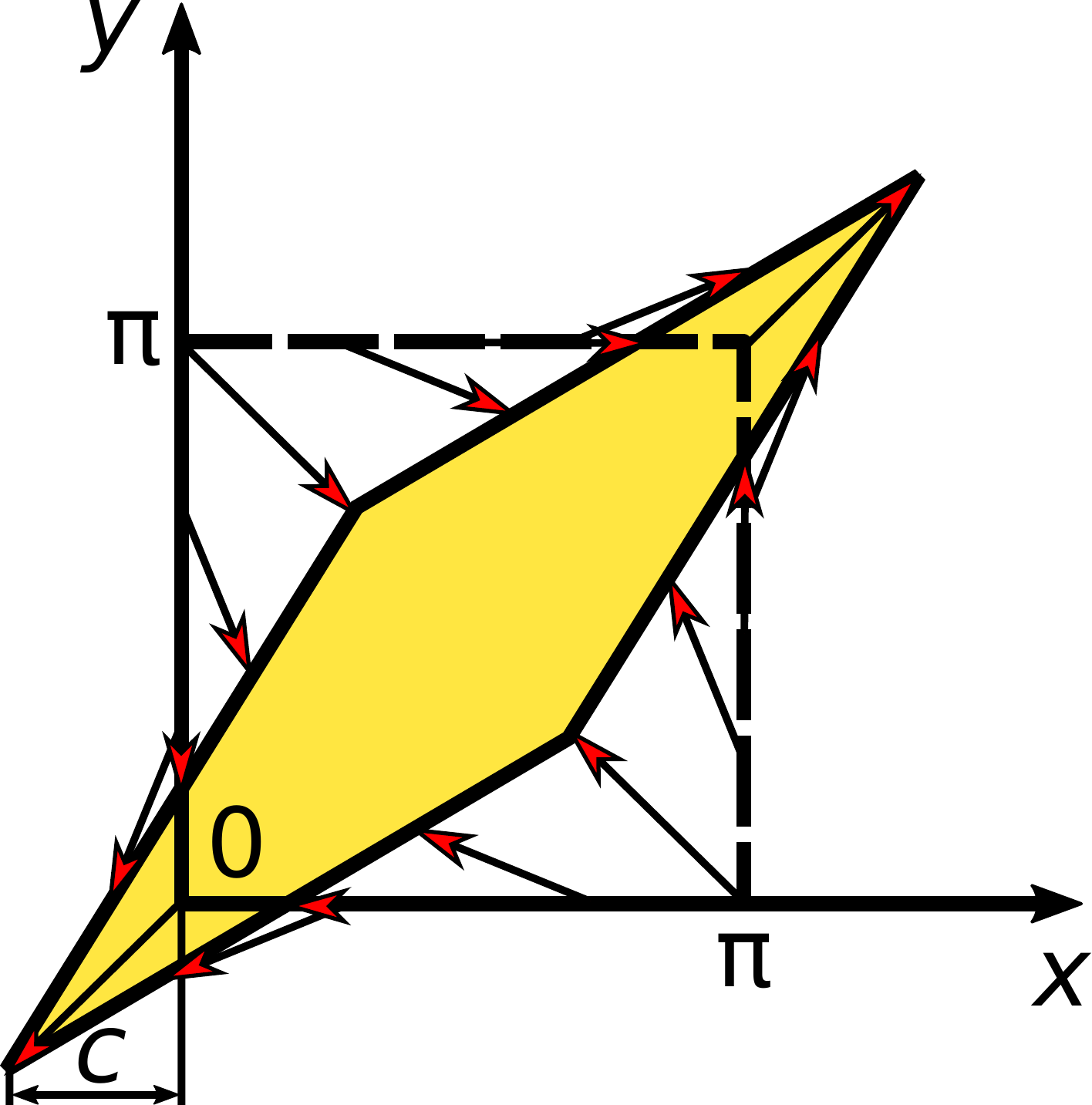}
\captionof{figure}{The sketch of the pure shear.}
\label{fig:sketch}
\end{figure}

One of the examined examples was pure shear. The boundary conditions are shown in table \ref{tab:purehf} and one can see the outline of the deformation in figure \ref{fig:sketch}.

\begin{table}[H]
\centering
\begin{tabular}{|c|c|c|}
\hline
boundary      & $u_x$         & $u_y$ \\ \hline \hline
$y=0$   & $-c$        &    $\frac{2c}{\pi}\left(x-\frac{\pi}{2}\right)$   \\ \hline
$x=\pi$ & $\frac{2c}{\pi}\left(y-\frac{\pi}{2}\right)$ &   $c$    \\ \hline
$y=\pi$    & $c$         &   $\frac{2c}{\pi}\left(x-\frac{\pi}{2}\right)$    \\ \hline
$x=0$   & $\frac{2c}{\pi}\left(y-\frac{\pi}{2}\right)$             &   $-c$    \\ \hline
\end{tabular}
\captionof{table}{The boundary conditions corresponding to pure shear, where $c$ is a constant that describes the extent of deformation.}
\label{tab:purehf}
\end{table}

\begin{figure}[H]
\centering
\includegraphics[width=11cm]{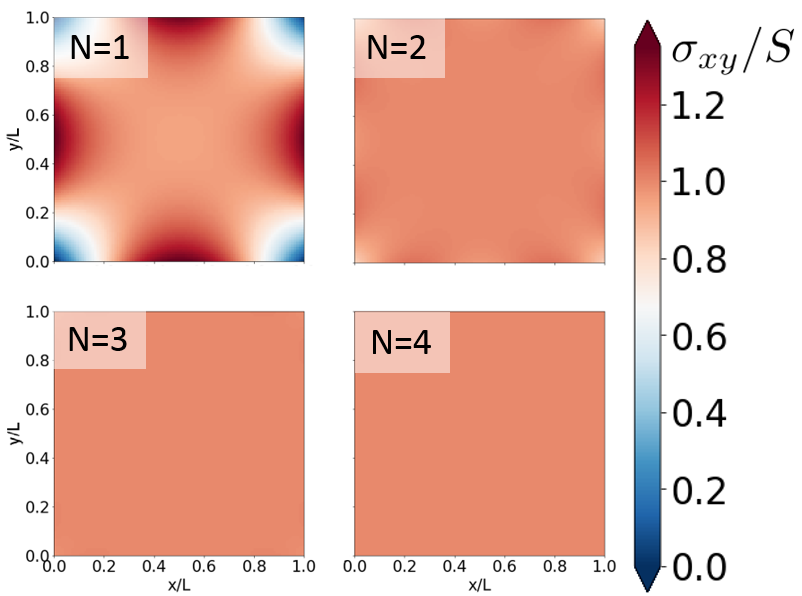}
\caption{The $\sigma_{xy}$ shear stress while applying pure shear. Here the shear modulus is $\mu=1$ and the extent of deformation is $c=1$ (for the meaning of $c$ see table $\ref{tab:purehf}$). Apparently, while increasing the $N$ basis size, the solution quickly converges to the analytical solution, which is spatially homogeneous in accordance with equation (\ref{eq:pure:sigma}).}
\label{fig:pureshear_sxy}
\end{figure}

The solution is the following:
\begin{equation}
\begin{gathered}
u_x=\frac{2c}{\pi}\left(y-\frac{\pi}{2}\right),\\
u_y=\frac{2c}{\pi}\left(x-\frac{\pi}{2}\right),
\end{gathered}
\end{equation}
which apparently fulfils the boundary conditions (table \ref{tab:purehf}) and the Navier--Cauchy equation (\ref{eq:navca}) as well. From this the stress is:
\begin{equation}
\begin{gathered}
\sigma_{xx}=0,\\
\sigma_{yy}=0,\\
\sigma_{xy}=\frac{4\mu c}{\pi}\equiv S,
\end{gathered}
\label{eq:pure:sigma}
\end{equation}
where $\mu$ is the shear modulus and $c$ is the constant that describes the extent of deformation. The $\sigma_{xy}$ stress component of the solution provided by our method can be seen in figure \ref{fig:pureshear_sxy} for the first four orders of basis size. As we can see, with increasing $N$ $\sigma_{xy}$ converges to a spatially homogeneous value of $S$ predicted by equations (\ref{eq:pure:sigma}). The other stress components (not shown here) demonstrated similar convergence to the expected zero value.

\begin{figure}[H]
\centering
\begin{tabular}{ccc}
\begin{subfigure}[b]{0.32\textwidth}
\includegraphics[height=\textwidth]{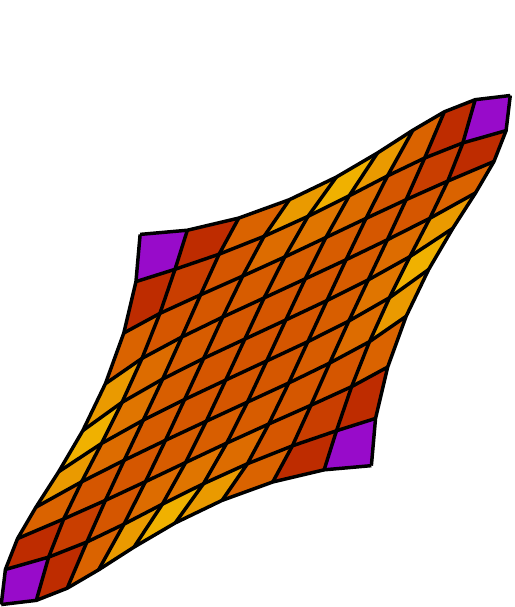}
\caption{$N=1$}
\end{subfigure} 
 &  
\begin{subfigure}[b]{0.32\textwidth}
\includegraphics[height=\textwidth]{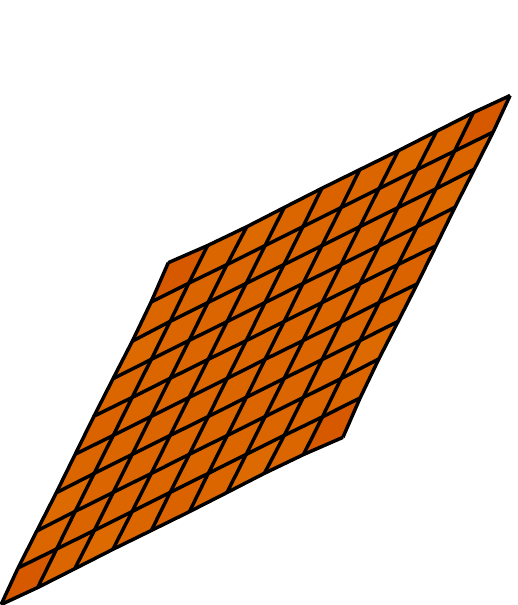}
\caption{$N=2$}
\end{subfigure} 
 &  
\begin{subfigure}[b]{0.32\textwidth}
\includegraphics[height=\textwidth]{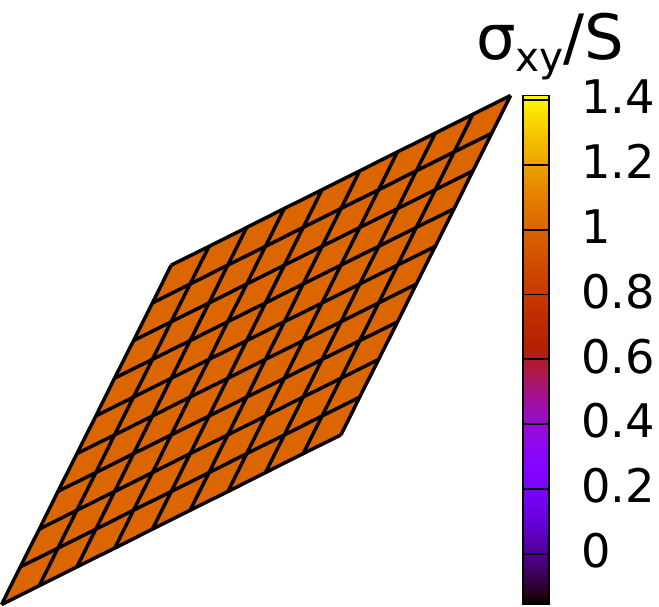}
\caption{$N=3$}
\end{subfigure}
\end{tabular}
\caption{The $\sigma_{xy}$ (colorbar) and the displacement of the pure shear for the first three orders of basis size.}
\label{fig:rombu}
\end{figure}

The displacement (and the $\sigma_{xy}$) corresponding to this deformation are shown in figure \ref{fig:rombu}. Both figure \ref{fig:pureshear_sxy} and figure \ref{fig:rombu} demonstrate that already the basis of $N=2$ describes the problem considerably well and much better than $N=1$. Since, $N=2$ is already very similar to the analytical solution, the improvement by further expansion of the basis is much smaller than from $N=1$ to $N=2$. Although, the convergence of the method seems obvious, we examined the convergence quantitatively.

The following quantities are introduced to characterize the deviation of the solution provided by our method from the analytical one.

\begin{gather}
p_1=\frac{1}{\mathcal{N}\sigma_{xy}^{t}}\sum_{i=1}^\mathcal{N} |\sigma_{xy}^{n}-\sigma_{xy}^{t}|,
\label{eq:dev1}
\\
p_2=\frac{1}{\sigma_{xy}^{t}}\sqrt{\frac{1}{\mathcal{N}}\sum_{i=1}^\mathcal{N} (\sigma_{xy}^{n}-\sigma_{xy}^{t})^2},
\label{eq:dev2}\\
p_\infty=\frac{1}{\sigma_{xy}^{t}}\mathrm{max}|\sigma_{xy}^{n}-\sigma_{xy}^{t}|.
\label{eq:devinf}
\end{gather}
Here $\mathcal{N}=11025$ is the number of points (placed on a square grid) where the analytical and numerical results of $\sigma_{xy}$ are compared. The upper 't' and 'n' indices denote the values provided by the theoretical solution and our numerical method, respectively. Obviously, these quantities could be calculated for other stress or displacement components as well.

\begin{figure}[H]
\centering
\begin{tabular}{cc}
\begin{subfigure}[b]{0.5\textwidth}
\includegraphics[width=\textwidth]{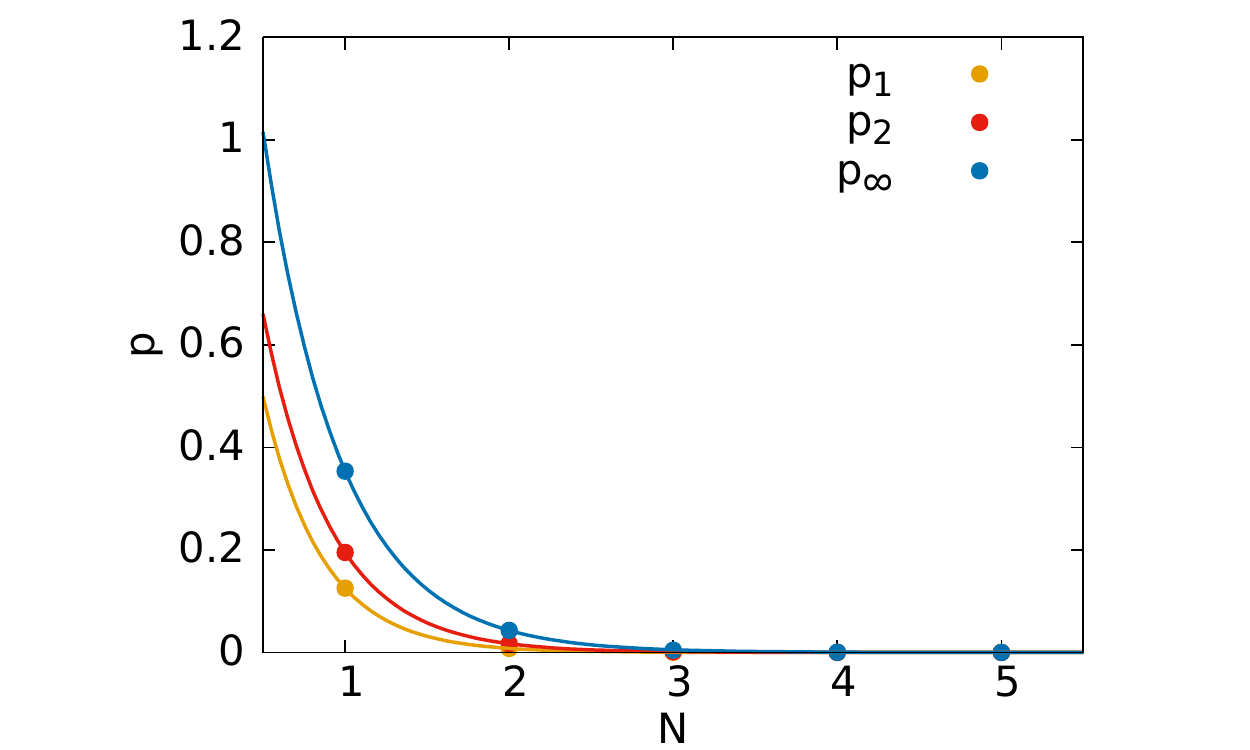}
\end{subfigure} 
 &  
\begin{subfigure}[b]{0.5\textwidth}
\includegraphics[width=\textwidth]{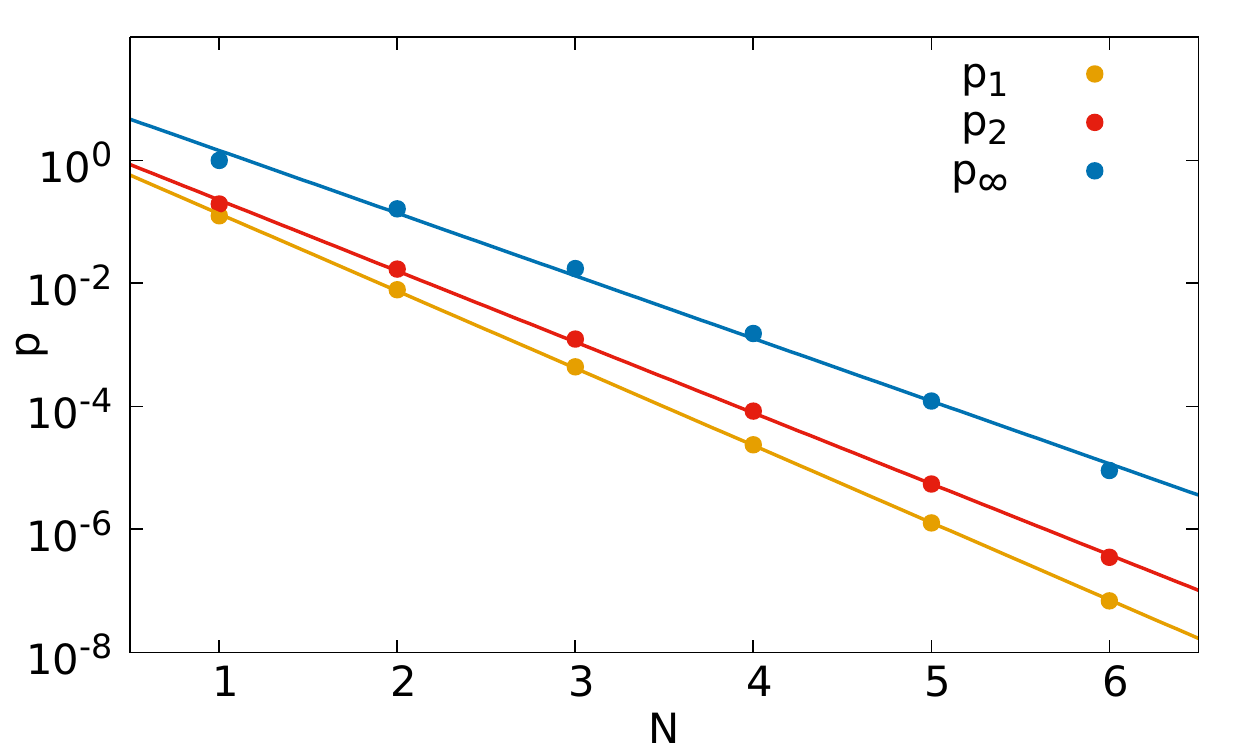}
\end{subfigure}    
\end{tabular}
\caption{The extent of deviation of $\sigma_{xy}$ from the analytical solution for pure shear on linear (left panel) and logarithmic (right panel) scales. For the definition of $p_1$, $p_2$ and $p_\infty$ see equations (\ref{eq:dev1}), (\ref{eq:dev2}) and (\ref{eq:devinf}).}
\label{fig:pure:error}
\end{figure}

As it is shown in figure \ref{fig:pure:error} all $p$ quantities decay quickly while increasing the size of the basis (i.e the value of $N$). The ($N$, $p$) points can be fitted well by the $Ae^{-bN}$ exponential function. The fit was done on ($N$, $\log{p}$) points using $\log{A}-bN$ trial function. The fit parameters and their standard deviation are shown in table \ref{tab:pure:fitparam}. This exponential convergence clearly outperforms the power-law convergence of FEM.

\begin{table}[H]
\centering
\begin{tabular}{|c|c|c|}
\hline
           & $A$             & $b$             \\ \hline\hline
$p_1$      & $2.36\pm0.14$ & $2.875\pm0.017$ \\ \hline
$p_2$      & $2.99\pm0.30$ & $2.627\pm0.031$ \\ \hline
$p_\infty$ & $4.11\pm0.60$ & $2.317\pm0.044$ \\ \hline
\end{tabular}
\caption{The parameters and their standard deviation of the $\log{p}=\log{A}-bN$ functions for $p_1$, $p_2$ and $p_\infty$ values (corresponding to $\sigma_{xy}$) for pure shear.}
\label{tab:pure:fitparam}
\end{table}

We investigated other analytically solvable problems ($\mathbf{u}=\mathrm{const}$; $u_x=cy$ and $u_y=0$) which showed similar fast exponential convergence with slightly different fit parameters. To summarize, the convergence properties of the numerical method (according to these analytically solvable problems) are promising since not only $p_1$ and $p_2$ but also $p_\infty$ showed fast exponential decay which means that there is no big deviation from the analytical solution. 

\subsection{Application on systems with dislocations}
\begin{figure}[H]
\centering
\includegraphics[width=0.9\textwidth]{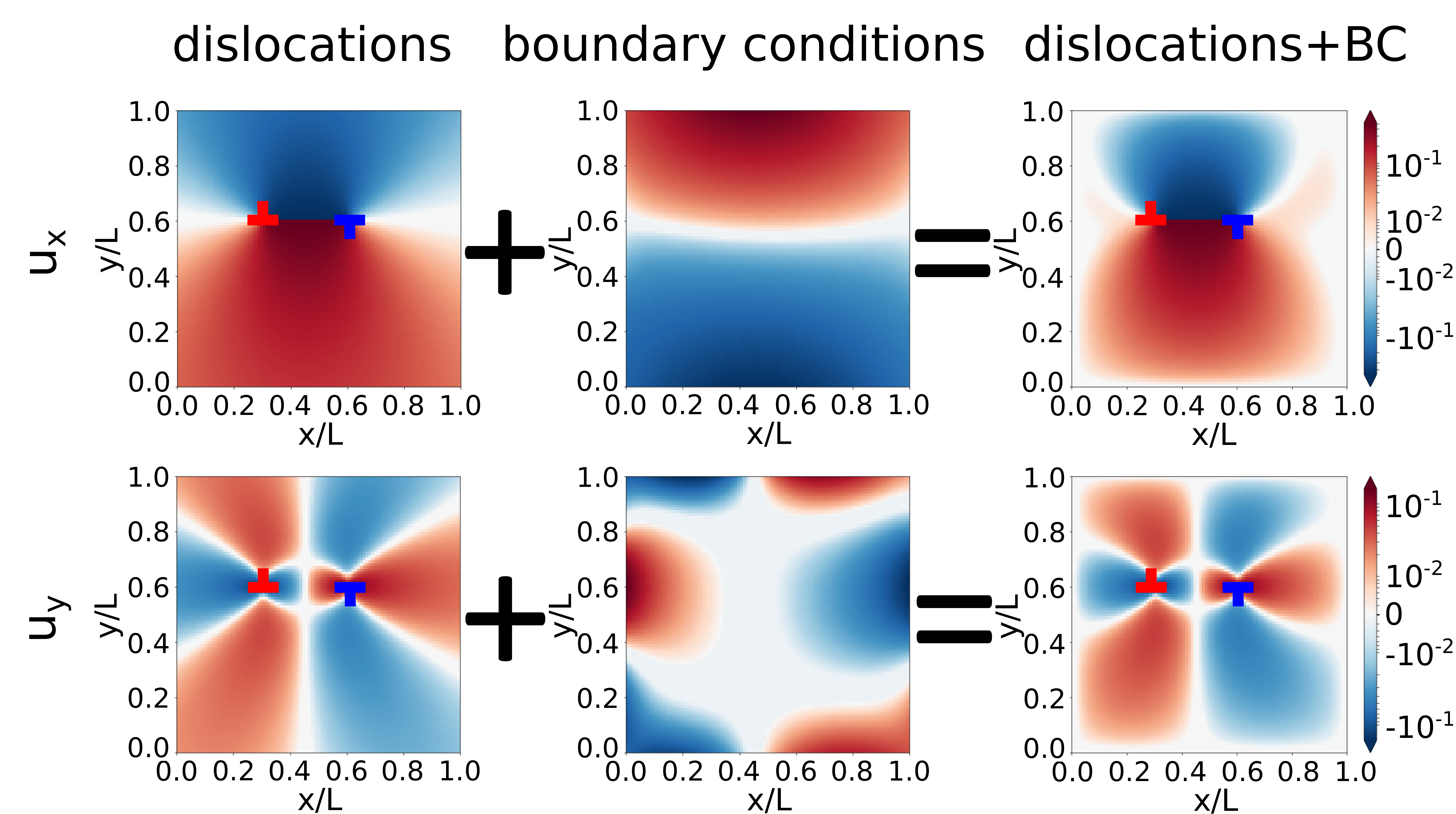}
\caption{The displacement field of a pair of dislocations with fixed boundaries ($\mathbf{u}|_\partial=0$). In this case we used $N=2$ basis size. Apparently, the method successfully creates the displacement field that contains the dislocations and fulfils the imposed boundary conditions.}
\label{fig:spu}
\end{figure}

As mentioned above we solve the dislocationless elastic problem to derive the solution containing dislocations using the principle of superposition (see figure \ref{fig:vdg}). For this purpose we determined the Fourier coefficients of the dislocations' (displacement or stress) field (valid in infinite medium) on the boundaries using FFT. The (finite number of) Fourier coefficients can be arranged in a vector $\mathbf{f_{\infty}}$. The boundary condition are also decomposed into Fourier series and the coefficients are settled in a vector $\mathbf{f_{BC}}$. The arrangement of Fourier coefficients in the vectors $\mathbf{f_{\infty}}$ and $\mathbf{f_{BC}}$ is arbitrary, although, should be the same in the two cases. The vector $\mathbf{f}$ from which the proper dislocationless field can be calculated (using the inverse of the matrix \textbf{M} ) is clearly the difference of these two vectors:
\begin{equation}
\mathbf{f}=\mathbf{f_{BC}}-\mathbf{f_\infty}.
\label{eq:bsup}
\end{equation}
As it was explained above, we can get the field of dislocations in finite system by adding their field to the appropriate dislocationless solution of the elastic problem derived from the vector \textbf{f}. The method is demonstrated in figure \ref{fig:spu} for the field of a dislocation dipole with Dirichlet boundary conditions, namely $\mathbf{u}|_\partial=0$. We tested on Neumann and mixed boundary conditions as well and found that the method can handle these two boundary conditions as well.

As a special case we investigated the force acting on a dislocation near an infinite free surface. The stress field of this system can be given analytically using a mirror dislocation \cite{hirth1992theory}. The force acting on the dislocation will be perpendicular to surface (due to symmetry reasons) and its magnitude is determined by the $\sigma_{xy}$ shear stress at the locus of the dislocation. The analytical solution yields that the magnitude of the force is
\begin{equation}
    F=\frac{\tau_0}{x}\propto \frac{1}{x},
\end{equation}
where $x$ is the distance of the dislocation from the surface and $\tau_0=-\frac{\mu b^2}{4\pi(1-\nu)}$ where $\mu$, $\nu$ and $b$ are the shear modulus, the Poisson ratio and the magnitude of the Burgers vector, respectively.

\begin{figure}[H]
\centering
\begin{tabular}{cc}
\begin{subfigure}[b]{0.45\textwidth}
\includegraphics[width=\textwidth]{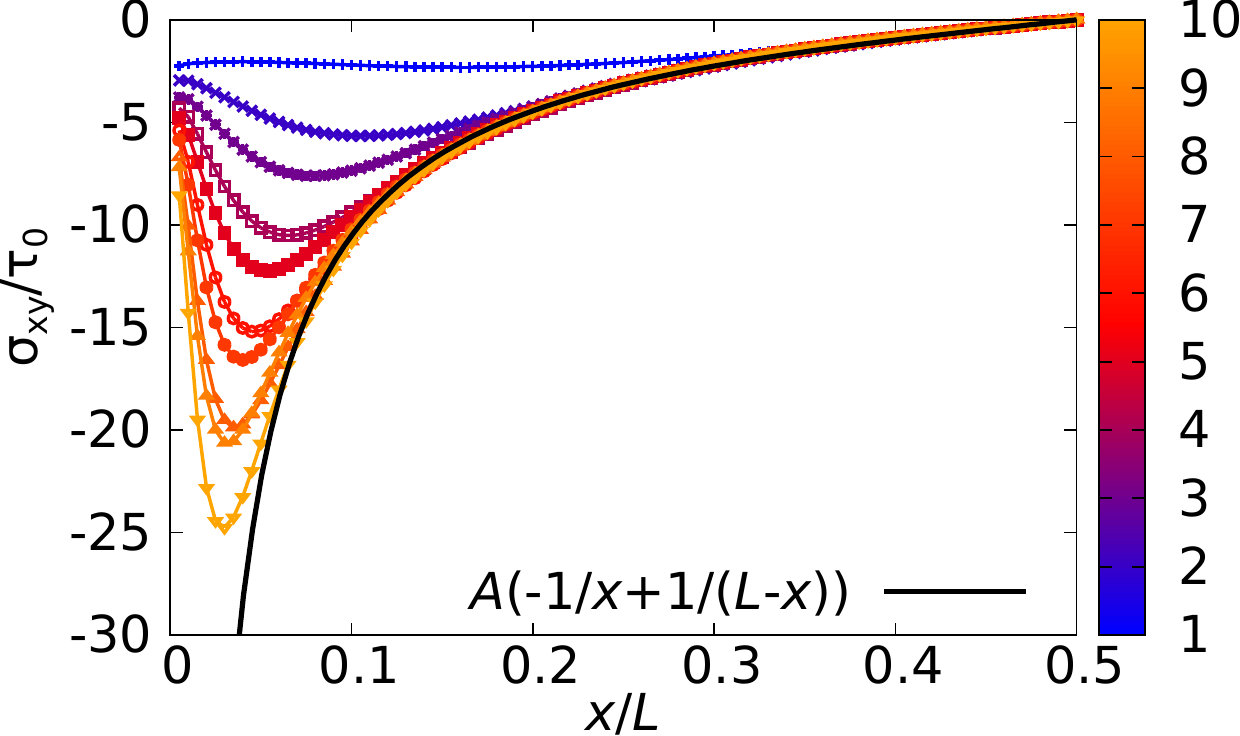}
\caption{}
\end{subfigure} 
 &  
\begin{subfigure}[b]{0.45\textwidth}
\includegraphics[width=\textwidth]{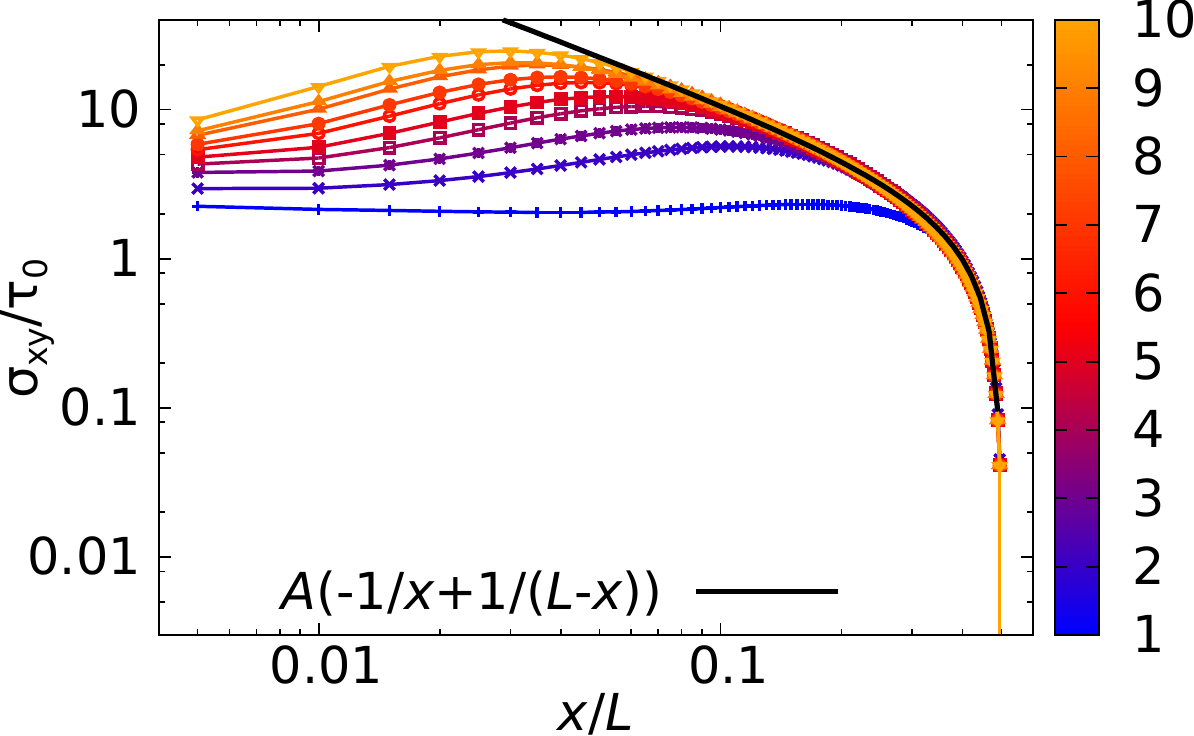}
\caption{}
\end{subfigure}    
\end{tabular}
\caption{The shear stress contribution of the boundary at the locus of the dislocation near free surface. The analytical $F=\frac{\tau_0}{x}\propto \frac{1}{x}$ dependence \cite{hirth1992theory} (where $\tau_0=-\frac{\mu b^2}{4\pi(1-\nu)}$) and the result provided by our numerical method for different $N$s are shown with lines of different colours. The fit curve is the superposition of the effect of two infinite free surfaces (at $x=0$ and $x=L$). The $A=1.18$ parameter was found to fit the data points. The deviation of $A$ from $1.0$ may be the consequence that theory assumes infinite boundary in the $y$ direction while we are working in a finite box.}
\label{fig:freesurf}
\end{figure}

A dislocation dipole was placed at ($x$,$L/2$) and ($L-x$,$L/2$) in the $L\times L$ rectangular 2D area. As the dislocations approach the boundaries, that is, $x$ tends to zero or $L$, we expect the results of the numerical method to approach the analytical solution, because in this case the effect of the nearest boundaries ($x=0$ and $x=L$) is much more significant than that of the farther ones and the other dislocation. We determined the external shear stress at the position of the dislocation for different $x$ values and different $N$ orders of calculation. We used $\nu=1/4$ during these calculations. In terms of boundary conditions the free surface yields $\sigma_{nn}=0$ and $\sigma_{xy}=0$ at the boundary where $n\in\{x,y\}$ is the direction perpendicular to the boundary. We specified this boundary condition on $x=0$ and $x=L$ boundaries. On the other two boundaries ($y=0$ and $y=L$) we prescribed that the displacement components of the image field must vanish. The results are shown in figure \ref{fig:freesurf}. It is consistent with the well-known fact that the free surface attracts the dislocation and the $\propto 1/x$ dependence is reproduced within a region in which the dislocation is not too far from the surface (hence the approximation of the infinite surface is valid) and not too close to it (where the numerical method does not work correctly due to the finite size of the basis). The width of the boundary region where the method does not work properly (where $F\propto 1/x$ is not met) decreases as $N$ increases. We note, however, that the sign of the field does not change in this boundary region, so it is (as physically expected) attractive in the whole vicinity of the surface.

It is important to investigate how the width of the boundary region (where the numerical results deviate from the analytical predictions) decreases with increasing $N$ to know what basis size should we use to achieve a certain desired precision. We defined a threshold where the slope (more precisely the discrete $\left| \frac{\Delta \log(-\sigma_{xy})}{\Delta (x/L)} \right|$ quotient) moves apart from the analytical value of $1$ and drops under an arbitrary $m$ value. 

\begin{figure}[H]
\centering
\includegraphics[width=\textwidth]{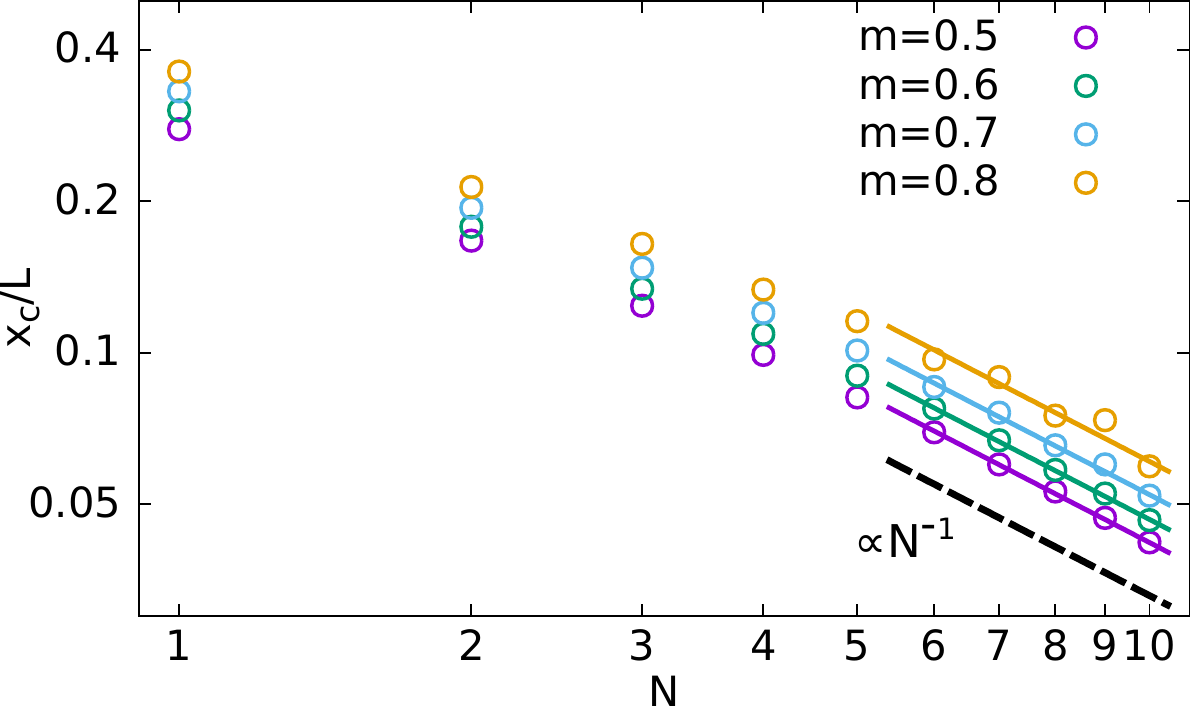}
\caption{The width of the boundary layer (where the model cannot reproduce the analytical diverging fields) for different slope thresholds $m$. The figure shows that independently of the value of parameter $m$, the data points show the expected $x_c/L\propto 1/N$ dependence.}
\label{fig:cutoff}
\end{figure}

As one can see in figure \ref{fig:cutoff}, the data showed $x_c/L\propto 1/N$ dependence for not too small $N$s.

Using the notation $\lambda_{\mathrm{min}}$ for the shortest wavelength occurring in the trigonometrical functions in the used basis functions, one gets that
\begin{equation}
    \lambda_{\mathrm{min}}\propto \frac{1}{N}.
\end{equation}
Also, the results above show that
\begin{equation}
    x_c\propto\frac{1}{N},
\end{equation}
yielding
\begin{equation}
    x_c\propto\lambda_{\mathrm{min}}.
\end{equation}
To sum up, the growth of the basis size $N$ reduces the width of the region near the surface where the method does not provide correct results. This width decreases proportionally with the shortest occurring wavelength $\lambda_{\mathrm{min}}$.

Now we turn, as a possible future application, to discrete dislocation dynamics simulations and pose the question how the basis size $N$ should be chosen in order to preserve precision over various system sizes (similar considerations apply for possible applications, such as CDD). Since, we are examining a 2D system, the characteristic distance $r_\mathrm{dis}$ between dislocations is proportional to $1/\sqrt{N_{\mathrm{dis}}}$, where $N_{\mathrm{dis}}$ is the number of dislocations. As we want to decrease the critical distance proportionally to this characteristic distance (i.e. $x_c\propto r_{\mathrm{dis}}$), the basis size $N$ should be increased proportionally to $\sqrt{N_{\mathrm{dis}}}$. It has important consequences on the computational time of the method as it will be discussed below. 

\subsection{Computational efficiency}
\label{subsec:compeff}
The computational cost is one of the most critical properties of numerical methods. In this section we examine the computational cost of each subtask that should be done once or after every time step in a DDD simulation. The results will be compared with the FEM and we found that our method has more favourable properties.

\begin{figure}[H]
\centering
\includegraphics[width=0.8\textwidth]{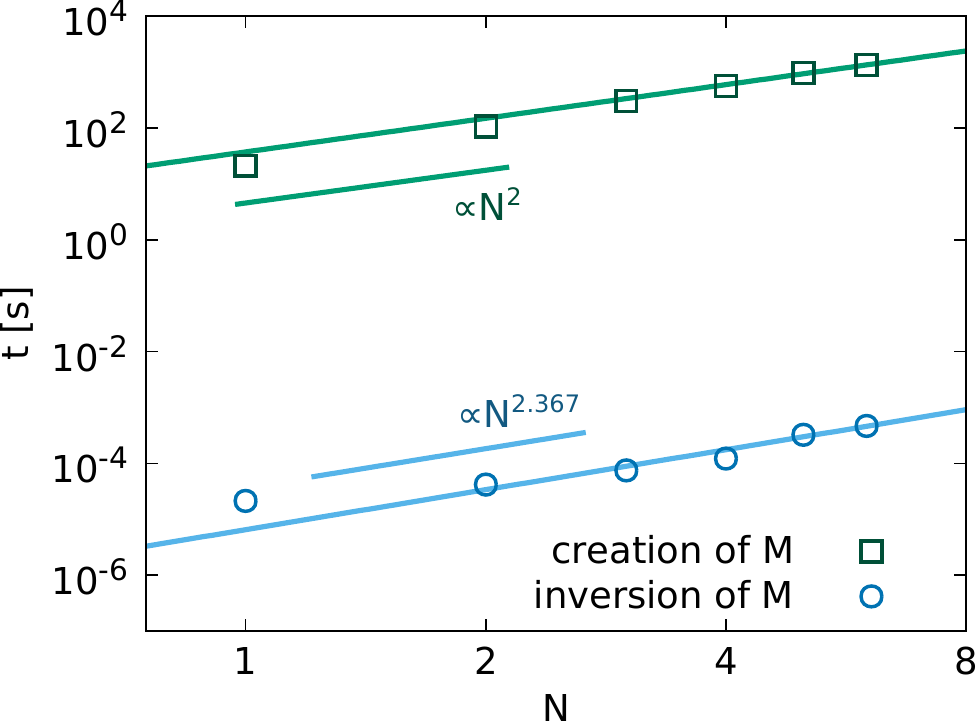}
\caption{The computational cost of non-repetable subtasks. The creation of the matrix \textbf{M} --with the size of $16N\times 16N$-- has $t_{\mathrm{M,create}}=c_1\cdot N^2$ computational cost, while its inversion costs $t_{\mathrm{M,inv}}=c_2\cdot N^{2.376}$. The value of the fit parameters is shown in table \ref{tab:timepara}.}
\label{fig:time:norep}
\end{figure}

If the type of boundary condition (Dirichlet, Neumann, etc.) is given, the matrix \textbf{M} remains the same (and so does its inverse) even if the concrete boundary values change. Hence, it is enough to evaluate and invert the matrix once. We found that the evaluation of the matrix has a computational cost of the form of $c_1 N^2$. It is plausible since all the $16N\times16N$ matrix elements should be calculated independently. Figure \ref{fig:time:norep} shows the computational cost of the construction and the inversion of the matrix \textbf{M} . We fit a $c_2 N^{2,376}$ function on the data points corresponding to the computational cost of matrix inversion with Coppersmith–Winograd algorithm \cite{coppersmith1990matrix}.

As a result of the motion of dislocations their field (valid in infinite medium) changes even on the boundaries, consequently, the $\mathbf{f_{\infty}}$ changes in time and so does the vector $\mathbf{f}$ -- according to Eq.~(\ref{eq:bsup}) -- even for unchanged boundary conditions (i.e.,~unchanged $\mathbf{f_{\mathrm{BC}}}$). To evaluate the vector $\mathbf{f}$ one needs to execute eight FFTs, since there are two relevant displacement or stress components on all four boundaries. Of course, if \textbf{f} changes, the vector \textbf{c} will do so as well, hence, we should perform the $\mathbf{c}=\mathbf{M}^{-1}\mathbf{f}$ multiplication every time step in a DDD simulation. To determine the motion of the dislocations one should evaluate the external stress (more precisely, the shear stress) at their locus caused by other dislocations as well. These three subtasks are all to be executed at every time step, therefore, their computational cost is critical in a DDD simulation.

\begin{figure}[H]
\centering
\includegraphics[width=0.8\textwidth]{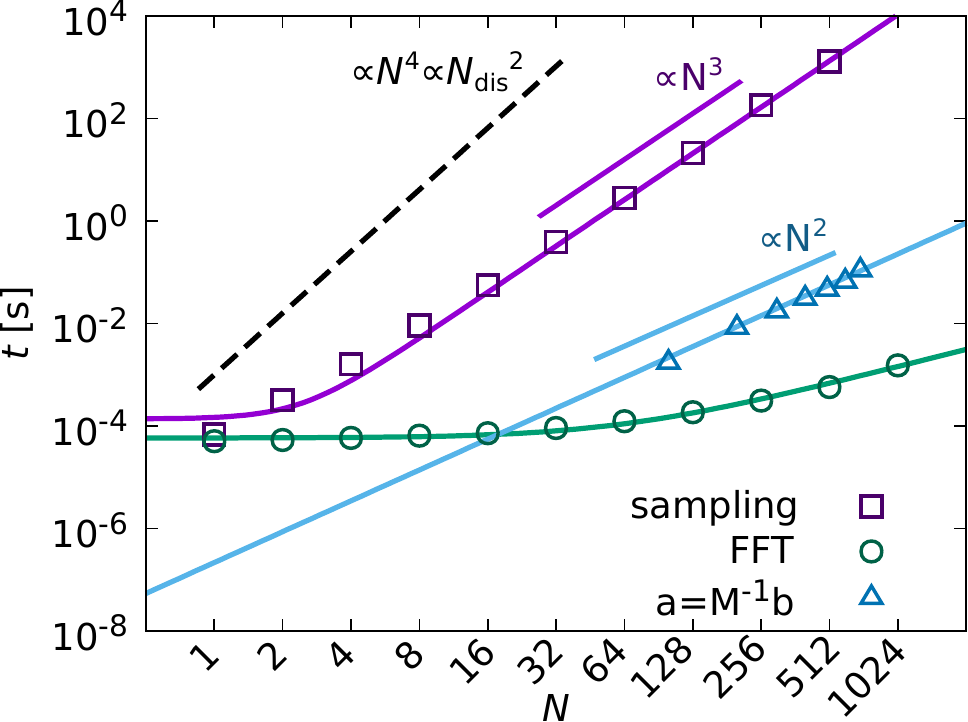}
\caption{The computational cost of repeatable subtasks. The computational cost of the evaluation of the field in $4N$ points on each boundary for $N_{\text{dis}}\propto N^2$ number of dislocations is $t_{\text{sample}}=c_3\cdot N^3 + c_4$. The execution of the FFT and the creation of the vector \textbf{f} has a $t_{\text{FFT}}=c_5 \cdot N \log(N)+c_6$ while the evaluation of the vector \textbf{c} has a $t_{a}=c_7 N^2$ time dependence. The latter is calculated from the $\mathbf{c}=\mathbf{M}^{-1}\mathbf{f}$ relation. The value of the fit parameters is shown in table \ref{tab:timepara}.}
\label{fig:time:rep}
\end{figure}

In order to execute FFTs along the boundary the field has to be evaluated at several points. Since we use only the first $2N$ Fourier-coefficients the number of these points is at least $4N$ on every boundary according the Nyquist--Shannon sampling theorem \cite{shannon1949communication}. The computational cost of this subtask is, therefore, $c_3 N N_{\mathrm{dis}}+c_4$, where $N_{\mathrm{dis}}$ is the number of dislocations since the field which is to be evaluated is the sum of the field of $N_{\mathrm{dis}}$ dislocations. The second subtask was to execute the FFT itself and settle the coefficients in the vector \textbf{f}. We assumed the computational time to be of the form of $c_5 N\log{N}+c_6$ where the first summand is the contribution of the FFT and the second one is the creation of the vector \textbf{f}. The last important subtask is the evaluation of the solution vector using the $\mathbf{c}=\mathbf{M}^{-1}\mathbf{f}$ relation. This is expected to have computational time of the form of $c_7 N^2$. The measured computational cost of these subtasks and the fits of the proposed functions can be seen in figure \ref{fig:time:rep}. The fit parameters are summarized in table \ref{tab:timepara}.

\begin{table}[H]
\centering
\begin{tabular}{|c|c|}
\hline
 $c_1$ & $(3.713\pm0.070)\cdot10^{1}$ s\\ \hline
 $c_2$ & $(6.53\pm0.32)\cdot10^{-6}$ s\\ \hline
 $c_3$ & $(9.861\pm0.076)\cdot10^{-6}$ s\\ \hline
 $c_4$ & $(6.9\pm2.9)\cdot10^{-5}$ s\\ \hline
 $c_5$ & $(1.968\pm0.066)\cdot10^{-7}$ s\\ \hline
 $c_6$ & $(5.9\pm1,6)\cdot10^{-5}$ s\\ \hline
 $c_7$ & $(2.177\pm0.080)\cdot10^{-7}$ s\\ \hline
\end{tabular}
\caption{The fit parameters of the functions fit on the measured computational cost of the relevant subtasks.}
\label{tab:timepara}
\end{table}

\section{Discussion}

In the following discussion we sum up our results about the computational efficiency of our numerical method and compare them with the properties of FEM which is often used to handle boundary conditions in DDD simulations.

In section \ref{subsec:compeff} we examined the region where the numerical method reproduces well the analytical solution for the force acting on a dislocation near an infinite free surface (see figure \ref{fig:cutoff}). Using these results we concluded that if the total number of dislocations $N_\text{dis}$ increases (thus, their characteristic distance decreases) one needs to increase the basis size $N$ in order to preserve the numerical precision as
\begin{equation}
    N\propto\sqrt{N_{\textrm{dis}}}.
    \label{eq:ndis:ours}
\end{equation}
The most time-consuming subtask that is to be executed every time step is the evaluation of the field at (at least) $4N$ points on the boundaries with a computational cost of $\mathcal{O}\left(NN_{\mathrm{dis}}\right)$ (see figure \ref{fig:time:rep}). Therefore, the leading-order term is
\begin{equation}
    t\propto NN_{\mathrm{dis}}\propto N_{\mathrm{dis}}^{3/2}
\end{equation}
where we utilized equation (\ref{eq:ndis:ours}).

While using FEM to solve PDE the problem leads to a system of linear equations. If it is described by an $N_\text{FEM}\times N_\text{FEM}$ matrix, the computational cost is $\mathcal{O}\left(N_\text{FEM}^2\right)$. The number of basis functions $N_\text{FEM}$ should be increased with the number of dislocations as
\begin{equation}
    N_{\mathrm{FEM}}\propto N_{\mathrm{dis}}
    \label{eq:ndis:fem}
\end{equation}
to get an sufficiently dense grid in a 2D simulation area. Hence, the computational time of the leading-order term reads as
\begin{equation}
    t_{\mathrm{FEM}}\propto N_{\mathrm{FEM}}^2\propto N_{\mathrm{dis}}^{2}.
\end{equation}

The results (expressed in terms of dislocation number $N_\text{dis}$) can be translated into the function of the linear size $L$ of the system if we assume a given dislocation density $\rho_{\mathrm{dis}}$. Then, in 2D
\begin{equation}
    N_{\mathrm{dis}}\propto\rho_{\mathrm{dis}}L^2\propto L^2.
    \label{eq:ndis:L}
\end{equation}
Our results discussed above (expressed with both $N_\mathrm{dis}$ and $L$) are summarized in table \ref{tab:timeconfront}.

\begin{table}[H]
\centering
\begin{tabular}{|l|c|c|}
\hline
 & $t(N_{\mathrm{dis}})$ & $t(L)$ \\ \hline \hline
our spectral method & $\mathcal{O}\left(N_{\mathrm{dis}}^{3/2}\right)$ & $\mathcal{O}\left(L^3\right)$ \\ \hline
finite element method & $\mathcal{O}\left(N_{\mathrm{dis}}^{2}\right)$ & $\mathcal{O}\left(L^4\right)$ \\ \hline
\end{tabular}
\caption{The computational cost of our method and the FEM expressed with the $N_{\mathrm{dis}}$ number of dislocations and the $L$ characteristic linear size of the system. Apparently, our method has more favorable properties, therefore, it can be more efficiently used to handle boundary conditions in DDD simulations.}
\label{tab:timeconfront}
\end{table}

The result shown in table \ref{tab:timeconfront} is remarkable, because simulating the interaction between dislocations has a computational cost of $\mathcal{O}\left(N_{\mathrm{dis}}^2\right)$. It is because there are $N_{\mathrm{dis}}(N_{\mathrm{dis}}-1)/2$ distinct pairs of them and due to the long-rangedness of dislocation stress fields all pair interactions have to be taken into account during the course of the simulation. Thus, our method (to handle boundary conditions) has more favorable computational complexity than the computation of interactions in DDD simulations (while FEM does not). So, the main point is that utilizing our method, taking the boundary conditions into consideration will not be the main component that limits the maximal number of dislocations or system size due to its good computational complexity, while the less favorable complexity of FEM can reduce its applicability.

\section{Summary}

In this paper a numerical spectral method has been proposed that provides a solution of the Navier—Cauchy equation (which describes homogeneous and isotropic medium) in 2D with given boundary conditions. The method is able to solve Dirichlet, Neumann and mixed boundary value problems as well. Since, the solution is a linear combination of basis functions which satisfy the equation exactly, in principle, it will also fulfil the equation exactly. However, the boundary conditions are only met approximately. The basis we use is finite, therefore, the possible solutions one can reproduce with this method are from a subspace of all solutions of the Navier—Cauchy equation. Thus, we had to find the approximate solution in this subspace that is the closest to the genuine solution in some sense. The proposed requirement is that the first finite number of Fourier coefficients of the Fourier series of the approximate solution on the boundary should be identical to the Fourier coefficients of the boundary condition.

Firstly, our method was tested on analytically solvable problems such as pure shear. The method reproduced the analytical solution and showed remarkable fast exponential convergence with the increment of the basis size which is superior to the power-law convergence of FEM. Secondly, the method was applied to cases where the simulation cell contained dislocation. It was found that if a dislocation is closer to the boundary than a certain distance (which decreases at higher orders of computation) numerical errors appear due to the analytically diverging stress fields. Based on this observation the time complexity that is needed to achieve a certain precision was assessed. As it was discussed in detail, the solution of the PDE leads to a $\mathbf{c}=\mathbf{M^{-1}f}$ type multiplication, where vector $\mathbf{f}$ can be obtained from the prescribed boundary values, matrix $\mathbf{M}$ is characteristic to the type of boundaries and vector $\mathbf{c}$ characterizes the solution function. In a typical application the matrix \textbf{M} is unchanged during a simulation even if the boundary values change (but remains of the same type, for instance Dirichlet), hence, it is enough to evaluate and invert the matrix once, while the  vector $\textbf{f}$ should be calculated at every time step. Naturally, the subtask that should be done every time step will be the ones that determine the computational efficiency of the method. After investigating the computational time of these subtasks we concluded that the computational complexity of our method is $\mathcal{O}\left(N_{\mathrm{dis}}^{3/2}\right)$, that is, $\mathcal{O}\left(L^3\right)$ where $N_{\mathrm{dis}}$ and $L$ are the total number of dislocations considered and the characteristic linear system size, respectively. Thus, contrary to FEM, the computational complexity of our numerical method is more favourable than the calculation of dislocation-dislocation interactions in DDD simulations (being $\mathcal{O}(N_\text{dis}^2)$ or, equivalently, $\mathcal{O}(L^4)$). Consequently, taking the boundary conditions into account will have a lower computational cost compared to other tasks, therefore, this component of the simulation will not limit the dislocation number (or system size) we are able to investigate in reasonable time. This allow us to examine larger systems and gain better statistics (of for example dislocation avalanches). In the future we intend to build in this method into already working DDD simulations that utilized PBC so far and investigate the effect of boundaries on dislocation avalanches and size effects. We also note, that the application of this method is not limited to DDD simulations, but its advantageous runtime properties can be also utilized in other 2D elastic problems such as in CDD.

\section{Acknowledgements}

We thank Géza Tichy for fruitful discussions. This work was completed in the ELTE Institutional Excellence Program (1783-3/2018/FEKUTSRAT) supported by the Hungarian Ministry of Human Capacities. The present work was supported by the National Research, Development and Innovation Fund of Hungary (contract numbers: NVKP\_16-1-2016-0014, NKFIH-K-119561, NKFIH-KH-125380). PDI is also supported by the ÚNKP-18-4 New National Excellence Program of the Hungarian Ministry of Human Capacities and by the János Bolyai Scholarship of the Hungarian Academy of Sciences.
\newpage

\addtocounter{section}{11}
\addcontentsline{toc}{section}{Hivatkozások}

\bibliographystyle{h-physrev_hu_notitle}
\bibliography{efficient}

\end{document}